# Eight orders of dynamical clusters and hard-spheres in the glass transition

Jialin Wu

College of Material Science and Engineering, Donghua University,

Shanghai, 200051, China, Email: jlwu@dhu.edu.cn

The nature may be disclosed that the glass transition is only determined by the intrinsic 8 orders of instant 2-*D* mosaic geometric structures, without any presupposition and relevant parameter. An interface excited state on the geometric structures comes from the additional Lindemann distance increment, which is a vector with 8 orders of relaxation times, 8 orders of additional restoring force moment (ARFM), quantized energy and extra volume. Each order of anharmonic ARFM gives rise to an additional position-asymmetry on a 2-*D* projection plane of a reference particle, thus, in removing additional position-asymmetry, the 8 orders of 2-*D* clusters and hard-spheres accompanied with the 4 excited interface relaxations of the reference particle have been illustrated. Dynamical behavior comes of the slow inverse energy cascade to generate 8 orders of clusters, to thaw a solid-domain, and the fast cascade to relax tension and rearrange structure. This model provides a unified mechanism to interpret hard-sphere, compacting cluster, free volume, cage, jamming behaviors, geometrical frustration, reptation, Ising model, breaking solid lattice, percolation, cooperative migration and orientation, critical entanglement chain length and structure rearrangements. It also directly deduces a series of quantitative values for the average energy of cooperative migration in one direction, localized energy independent of temperature and the activation energy to break solid lattice. In a flexible polymer system, there are all 320 different interface excited states that have the same quantized excited energy but different interaction times, relaxation times and phases. The quantized excited energy is about 6.4 $k$ = 0.55meV.

PACS numbers: 61.20.Gy, 63.70.+h, 64.70.Pf, 82.20.Rp,

## I. INTRODUCTION

One of the most challenging problems in condensed matter physics is to understand the glass transition [1, 2]. Recent theories of glass transition focus on the short-ranged attractive colloidal system [3-6], the Boson Peak [7-11], the mode-coupling theory [12, 13], the geometrical frustration [14], the jamming behavior [15, 16]. de Gennes [17] pointed out that the possible weakness of the mode-coupling theories is the ignorance of geometrical frustration effect and he set up a cluster picture with mosaic structure. The mosaic structure is based on the compact primary clusters which roughly correspond in size to the Boson peak [18]. Wolynes [9] proposed that quantized domain wall motions connected with the mosaic structure predicted by the random first-order transition theory [19]. More recently Wolynes shows that the transition states in a statistical field theory are instantons possessing additional replica symmetry breaking in the entropic droplet and argues that this new replica symmetry breaking state arises from configurational entropy fluctuations [20]. One of the most striking is that a population of fast particles moves co-operatively in string-like paths and clusters into transient, fractal structures that grow rapidly in size as the glass transition is approaching [21]. Weeks [22] emphasizes that the existence and behavior of the clusters indicate that the relaxations are very inhomogeneous, both temporally and spatially.

This paper mainly aims to find out the intrinsic mosaic geometric structures without any presupposition and relevant parameter in the ideal glass transition, from which current various theories and explanation on glass transition can be involuntarily deduced. The paper also tries to obtain a universal picture of cooperative migration of particle-clusters in a whole temperature range from $T_g$ to $T_m$. More complex jamming behavior of 200 chain-particle cooperative rearrangements will also be described in this paper. Before the hard-sphere model is modified, the current models and concepts in the glass transition have to be seriously considered as follows.

On the one hand, some experiments indicate a correlation between the nature of glass transition and the relative concentration of tunneling modes [23], which is a localized quantum mechanical two-level tunneling system. Although a broad range of experimental observations can be explained in terms of this idea there are many open questions concerning the standard two-level model [24]. The simple "two-state" models [25] have recently reappeared [26-29]. Wolynes [9] also have pointed out that the multilevel behavior of these domains can be pictured as involving the concomitant excitation of the *ripplon* modes of the glassy mosaic along with the transition between local minimum energy configurations of a mosaic cell. Thermally generated spontaneous density fluctuation gives rise to an interface tension. The interface tension acts as a restoring force on the fluctuation and it excites an interface tension wave, which is called a capillary wave or ripplon. The capillary waves occurring on liquid-liquid interfaces have quantized energy [9, 32] and various wavelengths, and each frequency depends on each wavelength. Especially, one of the important results proposed by Wolynes, the *l*-th ( $l = 2, …9$) mode of a sphere is $(2l + 1)$-fold degenerate and here are no harmonics of higher than 9th to 10th order, corresponding to an atomic scale half-wavelength. This result of existence of the limited domain wall vibration frequencies will be applied to form the interfaces of 8 orders of different transient 2-*D* clusters in size in this paper.

On the other hand, in polymer physics, macromolecules can only move extremely slowly in the topological direction of segment according to the reptation model (or the tube model) proposed by de Gennes [33]. The reptation model in melting state must arise from the solid-to-liquid glass transition. Thereby, in glass-thawing, the cluster migration should be to a certain extent restricted by reptation of a macromolecular chain. This hypothesis is conceivable that the glass transition has the molecular weight dependency [34]. Current hard-sphere models, however, have not been able to reflect such restrictions. It is noteworthy that Garrahan's [35] result and Wolynes's [30] picture of site-excitation also have more or less reflected such restrictions. These pictures are reminiscent of the spatial fluctuations of an order parameter close to a continuous phase transition, such as the magnetization of an Ising model near criticality, and the order parameter is purely a dynamic object [9, 36].

## II. CONCOMITANT 8 ORDERS OF TRANSIENT 2-DIMENSION CLUSTERS

In flexible polymer system, there are two well-known empirical laws: one is Williams–Landel-Ferry (WLF) equation [43] to describe temperature and time dependent behaviour in the glass transition, the other is the so-called macromolecular entanglement law, i.e. the viscosity depends on the molecular weight as $\eta \sim M^{3.4}$, independent of temperature, solvent and molecular species [44]. All the two empirical laws deal with the microscopic descriptions of viscosity in ideal random system, which includes the number of particles in cooperative migration, needful energy and time to excite particle-clusters *along one direction*. If the excited energy is quantized, as in the Ising model, a local z-axial direction can be selected as the migrating direction in a referenced 3-*D* local field (chain-particle $a_0$ as center in Figs.2, 3, 4). This chapter will detailed discuss the thawing processes in the referenced local z-component space, or say, on a local 2-*D* x-y projection plane for a domain.

### A. Excited interface

According to the reptation model [33], the thawed reference particle $a_0$ will migrate along the local z-axial, which is also the z-component direction of covalent bond (or strong bond in small molecular system) connecting with $a_0$. The condition of migration for $a_0$ is that the 4 Van der Waals' interfaces which encircle $a_0$ should be first thawed, or say, the 4 interfaces of $a_0$ should be first excited on $a_0$ local x-y projection plane. Thus, the concept of microscopic interface excited state should be introduced, which is also a vector with relaxation time spectrum, on $a_0$ local x-y projection plane.

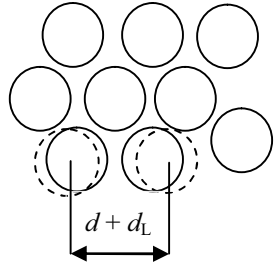

FIG. 1. The additional Lindemann distance increment $d_L$ between two hard-spheres (dashed line) appears by thermally excited ripplon in solid-to-liquid glass transition. $d_L \approx 0.10\sigma$ is a universal constant for all materials. $\sigma$, the diameter of hard-sphere.

In flexible polymer system, see Fig.1, when a thermally excited ripplon (vortical interface capillary wave) with a *vector component* appears on the interface between two neighboring reference particles $a_0$ and $b_0$, the additional Lindemann distance increment $d_L$ [9] giving rise to an *extra volume*, $\Delta v$, accompanying with the additional quantized excited energy $\Delta\varepsilon$ will occur on the interface. This interface is called the microscopic *excited interface* between $a_0$ and $b_0$ sites, and denoted as $j(a_0\uparrow b_0)$, see Fig.2. The arrow↑, from 1 to 2, represents the projection wave vector component between sites $a_0$ and $b_0$ on x-y projection plane of ripplon vector surrounding local z-axial. Note that the interface excitation $j(a_0\uparrow b_0)$ also comes from the cooperative contribution of the neighboring particle-sites $a_i$ and $b_i$ ($i \in [0, 8]$), see Fig.2, here comes the result of Wolynes: $j(a_0\uparrow b_0)$ only contains the contribution of 8 harmonic frequencies. The interface excitation thus has the relaxation time spectrum $\tau_i$ ($i \in [0, 8]$), so the $\Delta v$ and the $\Delta\varepsilon$ should be respectively modified as $\Delta v(\tau_i)$ and $\Delta\varepsilon(\tau_i)$. As the contribution of different neighboring particle-pairs at different times to the interface excitation between sites $a_0$ and $b_0$ always keep the direction of projection wave vector component, i.e., the direction of the arrow ↑ between sites $a_0$ and $b_0$, the energy quantization of interface excitation is hidden. That means *once the first excited interface with the direction 1 → 2 shown in Fig.2 appears in local field, the directions of all the succedent excited interfaces will be completely confirmed so as to form a interface excitation energy flow* as shown in Fig.3 - 6, otherwise, the total excitation energy of the system will increase. That is the reason and also the advantage to bring the signature of arrowhead into the standout mode of excited interface so as to get the minimum energy in exciting glass transition. Here the interface excitation represented by arrow is scaled by the quantized excited energy $\Delta\varepsilon(\tau_i)$ and thus the arrow is of a *unit length* for different wave lengths or frequencies on the lattice model. The convenience of this description is that the contribution of different frequencies (relaxation time) to interface excitation on $j(a_0\uparrow b_0)$ will be expressed by different sizes of 2-D symmetric loop flow surrounding $a_0$.

All interfaces in excited state with the extra volumes are correlative in space-time. The geometric method will be used in this paper to refract the interface-interface correlation in local space-time. Quantized excited energy $\Delta\varepsilon$ does not depend on temperature in flexible polymer system. The increase of temperature only *increases the number of interface excitation per unit time* to accelerate the thawing of domain frozen in the glass transition.

It is convenient that the model is described in terms of a lattice scaled by $\Delta\varepsilon(\tau_i)$. Whole interface excitations can be represented on the lattice formed by square cells because one chain-particle has 4 interfaces that can be excited. In Fig.2, the thick arrow 1-2 represents the direction of interface capillary wave with the extra volume $\Delta v(\tau_i)$ and excited energy $\Delta\varepsilon(\tau_i)$. Therefore, all energy (expressed by $kT$) to excite solid-to-liquid glass transition comes from the additive energy flow, in manner of flow-percolation formed by all excited interfaces in 3-D space. The number and the distribution of the interface excitations should meet the *minimum energy requirement* to excite glass transition. One of the singularities of glass transition is that the thermally excited ripplons only occur on the interfaces of a few thawed domains, which are able to form a energy flow-percolation during the observation time. That is consistent with the dynamic heterogeneity [27] and non-ergodic [37]. What is being cared is that (i) how much the number of whole excited interfaces is, i.e. the *number of whole microscopic excited states*, to thaw a domain in the glass transition, (ii) how much the flow-percolation energy is to form an *excited interface energy flow* in 3-D space.

### B. Additional restoring force moment

The position of particle $a_0$ or $b_0$ in solid state in Fig.2 is in equilibrium, thus the universal Lindemann displacement $d_L$ in Fig.1 generates a pair of additional restoring force moments (ARFMs), or called as torque, denoted as $M_0(a_0\uparrow b_0)$, which is the contribution coming from the interface tension between the pair of particles $a_0$ and $b_0$, or the torque with relaxation time $\tau_0$ between the two "solid-state" cells of $a_0$ and $b_0$. $M_0(a_0\uparrow b_0)$ also gives rise to an additional position-asymmetry on the projection plane. There are 8 orders of ARFMs (8 orders of torques with $\tau_i$) on $j(a_0\uparrow b_0)$, which come from the contributions of the pair of microscopic "solid state" blocks of $(a_0 + a_1 \ldots + a_i)$ and $(b_0 + b_1 \ldots + b_i)$ in Fig.2, denoted as $M_i(a_0\uparrow b_0)$, in which torque energies are also stored in the excited interface $j(a_0\uparrow b_0)$. Here the theoretical treatment is made to the transient lattice model: the harmonics wavelength in ripplon corresponds to the length of such "solid-state" lattice-block, whereas the homologous anharmonicity in ripplon is represented by both the behavior of mosaic-blocks with inlaid characters and the 8 orders of compacted 3-D clusters with more complex mosaic behaviors. A region of space that can be identified by a single mean field solution is called a mosaic cell [38]. The details of additional replica symmetry breaking (ARSB) using geometric method are going to be discussed in the following part.

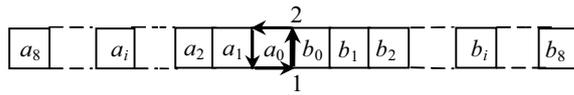

FIG. 2. Sketch of the interface excitation between two sites $a_0$ and $b_0$ on x-y projection plane in $a_0$ local field. The thick arrow $\uparrow$ represents the projection direction of interface capillary wave. Each arrow is scaled by the quantized interface excited energy $\Delta\varepsilon(\tau_i)$ and possesses a unit length for different relaxation times.

### C. Definition of excited particle by 4 excited interfaces

In the denotation of $j(a_0\uparrow b_0)$, it is denoted that the particle on the left side of the arrow $\uparrow$ in the parentheses (in this case $a_0$ in Fig. 2) will be *first* (because of phase difference of $\pi$ between $V_0(a_0\uparrow b_0)$ and $V_0(b_0\uparrow a_0)$) surrounded by 4 one after another excited interfaces to form a minimum loop (in the form of excited interface energy flow), as the 4 arrows in Fig.2 or the 4 thick arrows in Fig.3 (b), denoted as $V_0(a_0\uparrow b_0)$. $V_0(a_0\uparrow b_0)$ defines a particle $a_0$ surrounded by z-components of interface capillary waves in solid-liquid glass transition and has topological properties: cyclic direction, cyclic jumping-off point and the first interacting interface $j(a_0\uparrow b_0)$ in the 4 neighboring interfaces of $a_0$. Clearly, there is a phase difference of $\pi$ between $V_0(a_0\uparrow b_0)$ and $V_0(b_0\uparrow a_0)$ and the latter denotes the excited particle $b_0$ in $b_0$ (local) field. The denotation of the excited particle $V_0(a_0\uparrow b_0)$ can be simplified as $V_0(a_0)$.

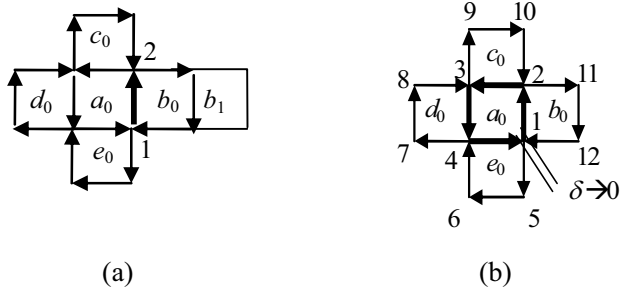

(a)　　　　　　　　(b)

FIG. 3. (a) The first order additional restoring force moments $M_1(a_0\uparrow b_0)$ (thick arrow 1-2) breaks the position-symmetry in loop $V_0(a_0\uparrow b_0)$ of particle $V_0(a_0)$ and new additional removing asymmetrical process should be adopted in the succedent evolution. (b) The symmetric $V_0(a_0)$ interacts one by one with its four symmetric neighboring excited particles, $V_0(e_0)$, $V_0(d_0)$, $V_0(c_0)$ and $V_0(b_0)$, and the 12 excited interfaces (thin arrows) form $V_1(a_0\uparrow b_0)$, which is the first order of transient symmetrical loop of the reference particle $V_0(a_0)$ in $a_0$ local field. Within $V_1(a_0\uparrow b_0)$, the interface tension relaxation of $\tau \leq \tau_1$ has been realized on the 4 interfaces of $V_0(a_0)$.

Within the loop of $V_0(a_0\uparrow b_0)$, the Lindemann displacements of the 4 excited interfaces let particle $a_0$ replace to the *primal equilibrium position* in real space, while to the central of block $a_0$ in the (energy) lattice model.

### D. First order 2-D symmetric loop and first order transient 2-D cluster

As the interface tension with $\tau_0$ on $j(a_0\uparrow b_0)$ comes from the contribution of the torque between two particles $a_0$ and $b_0$, thus the condition to form symmetric cycle of $V_0(a_0\uparrow b_0)$ for eliminating additional position-asymmetry is that the 4 neighboring symmetric loop of $a_0$: $V_0(b_0\uparrow a_0)$, $V_0(c_0\uparrow a_0)$, $V_0(d_0\uparrow a_0)$ and $V_0(e_0\uparrow a_0)$, should also be one by one finished (because of the Lindemann displacement $d_L$ being the *coupling value* of two particles). At the instant time $t_1$ in $a_0$ (local) field, once 4th cycle $V_0(e_0\uparrow a_0)$ is finished, a new loop of $a_0$ surrounded by the excited interfaces with $\tau_1$ timescale will occur as in Fig. 3(b). The new symmetric loop flow is defined as the interfaces of the *first order of transient 2-D cluster* in $a_0$ local field, which is denoted as $V_1(a_0\uparrow b_0)$ and simplified as $V_1(a_0)$. The cycle path of $V_1(a_0\uparrow b_0)$ is 1→ 5→ 6→ 4→ 7→ 8→ 3→ 9→ 10→ 2→ 11→ 12→1, whose cycle direction is negative, contrary to that of $V_0(a_0\uparrow b_0)$. $V_1(a_0\uparrow b_0)$ turns out to be the first order of concomitant $2\pi$ closed loop with particle $V_0(a_0)$ at the instant time $t_1$ in $a_0$ local field and at the moment the additional position-asymmetry of particle $V_0(a_0)$ has been eliminated within $V_1(a_0)$ and the interface tension relaxations of $\tau \leq \tau_1$ have been realized on the 4 excited interfaces of $V_0(a_0)$.

First order symmetry loop has 12 excited interfaces and is of the interface energy of $12\Delta\varepsilon$ ($\tau \leq \tau_1$). The number of cooperative excited particles in first order symmetry loop is 5 (i.e. $a_0$, $b_0$, $c_0$, $d_0$ and $e_0$ in Fig.3 (a)). Forming a first order symmetry loop deals with the *determinate path* of the 16 'self-avoiding walks' of excited interfaces, i.e., the given 16 interfaces to form $V_0(a_0\uparrow b_0) + V_1(a_0\uparrow b_0)$ illustrated in Fig.3 (b). The walk path of the given 16 self-avoiding walks in first order symmetry loop flow is the elementary method to form the following orders of 2-D clusters to ARSB in larger areas.

### E. Second order of transient 2-D cluster

Once $2\pi$ cycle $V_1(a_0\uparrow b_0)$ is finished, the reference particle $V_0(a_0)$ will replace the center of block $a_0$, whereas ARFMs $M_2(a_0\uparrow e_0)$ with $\tau_2$ on $j(a_0\uparrow e_0)$ will first and again break the position-symmetry of particle $V_0(a_0)$ in the block $a_0$, as in Fig. 4. For example, ARFMs with $\tau_2$ on $j(a_0\uparrow e_0)$ can be regarded as the contribution from the two "blocks" of $(a_0 + c_0 + c_1)$ and $(e_0 + e_1 + e_2)$ in Fig.4, thus a new process to eliminate position-asymmetry should be again adopted in the succedent evolution. In order that the reference particle $V_0(a_0)$ again replaces the center of the block $a_0$, 4 symmetric loops: $V_1(c_0\uparrow a_0)$ in $c_0$ field, $V_1(d_0\uparrow a_0)$ in $d_0$ field, $V_1(e_0\uparrow a_0)$ in $e_0$ field and $V_1(b_0\uparrow a_0)$ in $b_0$ field, should one by one be adopted in $a_0$ field. Therefore, the 2nd order of transient 2-D cluster with $\tau \leq \tau_2$, denoted as $V_2(a_0\uparrow b_0)$, is formed at the instant time $t_2$ when the 2nd order $2\pi$ loop finished in $a_0$ field, whose cycle direction is positive, contrary to that of $V_1(a_0\uparrow b_0)$. Note that the "solid-block" $e_2$ in the torque $M_2(a_0\uparrow e_0)$ on $j(a_0\uparrow e_0)$ in Fig. 4 is a mosaic cell that does not appear in the succedent $V_2(a_0\uparrow b_0)$.

The number of interfaces of $V_2(a_0\uparrow b_0)$ loop is 20 (20 thick arrows in Fig.4). The number of interfaces of torque relaxed with $\tau \leq \tau_2$ inside $V_2(a_0\uparrow b_0)$ loop is 12, and equals to the number of the interfaces of $V_1(a_0\uparrow b_0)$. This means that the evolution energy from $V_1(a_0\uparrow b_0)$ to $V_2(a_0\uparrow b_0)$ is $8\Delta\varepsilon$ ($\tau \leq \tau_2$). As same as generating $V_1(a_0\uparrow b_0)$, forming $V_2(a_0\uparrow b_0)$ in $a_0$ field is the contribution of 4 $V_1$-clusters in 4 neighboring local fields of $a_0$. This evolution rule holds true for the concomitant 8 orders of transient 2-D clusters in $a_0$ field. The number of cooperative excited particles in second order of cluster is 13 (13 instant cells (small squares) surrounded by 20 excited interfaces in Fig.4).

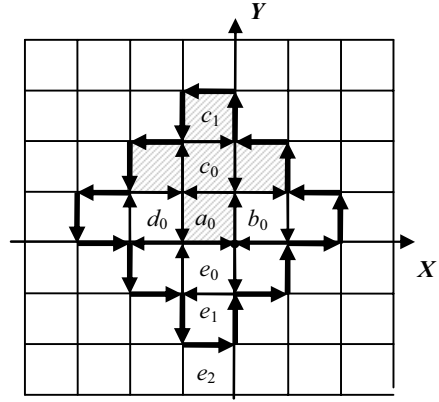

FIG. 4. Second order of transient 2-D cluster in $a_0$ local field, $V_2(a_0\uparrow b_0)$, only appears on the instant lattice graph at the instant time $t_2$. The shadow represents symmetric $V_1(c_0\uparrow a_0)$ loop in $c_0$ local field. Neighboring 4 symmetric loops, $V_1(c_0\uparrow a_0)$, $V_1(d_0\uparrow a_0)$, $V_1(e_0\uparrow a_0)$ and $V_1(b_0\uparrow a_0)$ cooperatively generate $V_2(a_0\uparrow b_0)$ in $a_0$ local field. The "solid-block" $e_2$ in the torque $M_2(a_0\uparrow e_0)$ on $j(a_0\uparrow e_0)$ is a mosaic cell that does not appear in the succedent $V_2(a_0\uparrow b_0)$.

### F. Third order of transient 2-D cluster

In the same way, the concomitant third order of transient 2-D cluster at the instant time $t_3$ in $a_0$ local field, $V_3(a_0\uparrow b_0)$ can be obtained as in Fig. 5.

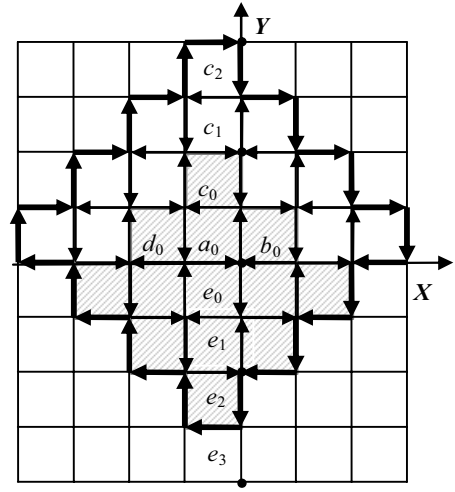

FIG. 5. The concomitant third order transient 2-D cluster $V_3(a_0\uparrow b_0)$ with the interface tension relaxations of $V_0(a_0)$ in the range of $\tau \leq \tau_3$. The shadow area represents the symmetric $V_2(e_0\uparrow a_0)$ loop in $e_0$ local field, whose symmetric center is $e_0$. The "solid-block" $e_3$ in the torque $M_3(a_0\uparrow e_0)$ on $j(a_0\uparrow e_0)$ in $a_0$ local field is a mosaic cell of $V_3(a_0\uparrow b_0)$ and appears with $V_4(a_0\uparrow b_0)$.

The number of interfaces of $V_3(a_0\uparrow b_0)$ loop is 28 (28 thick arrows in Fig.5). The number of interfaces of torque relaxed in the range of $\tau \leq \tau_3$ inside $V_3(a_0\uparrow b_0)$ loop is 20, which equals to the number of the interfaces of $V_2(a_0\uparrow b_0)$. This means that the evolving energy from $V_2(a_0\uparrow b_0)$ to $V_3(a_0\uparrow b_0)$ is 8 $\Delta\varepsilon$ ($\tau \leq \tau_3$). The number of cooperative excited particles in third order of cluster is 25 (25 instant cells surrounded by 28 excited interfaces in Fig.5), or say, 28 excited interfaces can excite 25 particles to cooperative thaw in third order of cluster. From the results of $V_1$ and $V_2$, it is clear that the more the number of cooperatively thawed particles is, the less the average needful excited interface energy for each particle is. The key point is that what the minimum excited energy is and what the number of cooperatively thawed particles excited by the energy in the glass transition is.

### G. 5-particle cooperative self-avoiding walk excited field

The number of interfaces of $i$-th order of transient 2-D cluster can be calculated as

$$L_i = 4(2i+1) = 12, 20, 28, 36, 44, 52, 60, 68 \text{ for } i = 1, 2\ldots 8 \tag{1}$$

Note that $L_8$ in eq (8) will be corrected as 60 by percolation, seeing following H section. The number of cooperative excited particles in $i$-th order of cluster is respectively:

$$N_i = N_{i-1} + 4i = 5, 13, 25, 41, 61, 85, 113, 145 \text{ for } i = 1, 2\ldots 8 \tag{2}$$

And $N_0 = 1$, in Figs.3 - 6. Note that $N_8$ also will be corrected as 136 by percolation, seeing H and III.B sections.

There are *4 neighboring concomitant excitation centers surrounding the referenced particle center.* Thus, this excited field can be also called 5-particle (cooperative self-avoiding walk) excited field to break solid-lattice. The forming of $i$-th order of cluster always results from the cooperative contributions of the 4 neighboring ($i$-1)-th order of clusters around the reference particle. During the time of ($t_{i-1}$, $t_i$), the 4 neighboring local coordinates can take any direction (fragmentized and atactic lattices) because of fluctuation. However, *at the instant time $t_i$,* the direction of $i$-th order cluster always starts *sticking to the direction* of the referenced first order cluster and forming the instant inerratic energy-lattices, such as Figs.3 – 6, to satisfy the *minimum energy requirement* of the glass transition. The energy to excite *i*-th order cluster orientable evolution thus is the energy of *one outer degree of freedom* with $\tau \leq \tau_i$ of cluster.

Each order of cluster has 8 more excited interfaces on x-y projection plane or in local z-component space than the cluster of the sub order. This means that the energy of one outer degrees of freedom to excite *i*-th order of cluster migrating along z-axial is equal to the evolution energy of $8\Delta\varepsilon(\tau_i)$ on x-y projection plane, which has no option but to equal the potential well energy $\varepsilon_i$ on z-axial of *i*-th order of cluster because generating a bigger cluster *always comes back to its equilibrium position on x-y projection plane* of the referenced $a_0$ particle. For ideal solid system, three potential well energies on x, y, z-axial always are equal. Therefore, the potential well energy in z-axial

$$\varepsilon_0(\tau_i) = 8\Delta\varepsilon(\tau_i), \tag{3}$$

It should be admitted that the energy $\varepsilon_0(\tau_i)$ in eq (3) in fact is the non-integrable geometric phase potential in multiparticle system. The interface excitations on a $2\pi$ closed loop belong to 'parallel transport' surround z-axial in topology analysis and the non-integrable geometric phase potential $\varepsilon_0(\tau_i)$ is $8\Delta\varepsilon(\tau_i)$ directly based on the definition of geometric phase [49].

$\varepsilon_0$ denoted as $\varepsilon_0(\tau_i)$, is a invariable value for flexible polymer (this result is as same as [39]), and $\varepsilon_i = 8\Delta\varepsilon_i$ ($\varepsilon_i \neq \varepsilon_{i+1}$) is of 8 orders of potential well energies for really complicated system. The latter may correspond to the so-called potential energy landscape [9, 39].

Each neighboring (*i*-1)-th order of cluster also has one outer degrees of freedom with $\tau \leq \tau_{i-1}$, so the *i*-th order of cluster in the referenced excited field is of 5 *inner degrees of freedom* with $\tau \leq \tau_i$ and taking any directions during the time of $(t_i, t_{i+1})$. Note that each excited particle on the x-y projection plane, in Fig.3 – 6, also connects with a covalent bond in excited state with excited energy $\Delta\varepsilon$, because of the resonance mode in critical phase transition, which can be in any direction. This also shows a picture of perfect random motion inside *i*-th order of cluster. Each order of 2-*D* cluster connects with 4 "unexcited solid state" mosaic cells that belong to the senior order of cluster.

### H. Eighth order of transient 2-*D* cluster

In the same way, the 8th order of transient 2-*D* cluster, $V_8(a_0\uparrow b_0)$, can be obtained in $a_0$ local field, as in Fig. 6. As there is no 9th cluster in system, percolation should be adopted at the instant time $t_8$, and the 8th order of cluster must be corrected for the anharmonicity in ripplon.

The uncorrected number of interfaces of $V_8(a_0\uparrow b_0)$ is 68, denoted by thick arrows in Fig. 6. When percolation appears in system, the 4 excited cells (with $\tau \leq \tau_8$) in $V_8(a_0\uparrow b_0)$, by thick upward diagonal lines, are the mosaic cells of in the 4 neighboring $V_7$-clusters, and 4 cells with $\tau \leq \tau_7$ in the 4 neighboring $V_7$-clusters, by thin upward diagonal lines, are the mosaic cells in $V_8(a_0\uparrow b_0)$. The 4 thick inverted arrows in Fig. 6 are the *mutual interfaces* of $V_8(a_0\uparrow b_0)$ and its 4 neighboring $V_7$-clusters. Thus the number of interfaces of $V_8(a_0\uparrow b_0)$ should be corrected as 60, i.e. $L_8$ (corrected by percolation) = 60

$$L_8 \text{ (corrected by percolation)} = 60 \tag{4}$$

Note that there are 4 *inverted* thick arrows in the corrected $V_8$ loop in Fig. 6. Thus, the number of concomitant excited particles with $V_0(a_0)$ in $V_8(a_0\uparrow b_0)$ should be corrected as 141 because the 4 cells (4 mosaic cells of $V_8$), marked by thin upward diagonal lines, belong to that of the neighboring $V_7$-clusters. In addition, the central 5 empty cells in $V_8(a_0\uparrow b_0)$ representing the interface tensions of 5 particles in $V_1(a_0\uparrow b_0)$ will be fully relaxed (i.e. the 5 particles will be fully in non-interacting with other 136 in 141 particles) when percolation appears. The 5 particles will first jump out of the projection plane of $V_8(a_0)$ after $a_0$ local time $t_8$. Therefore, the number of cooperative excited particles in 8th order of cluster should finally be corrected as 136.

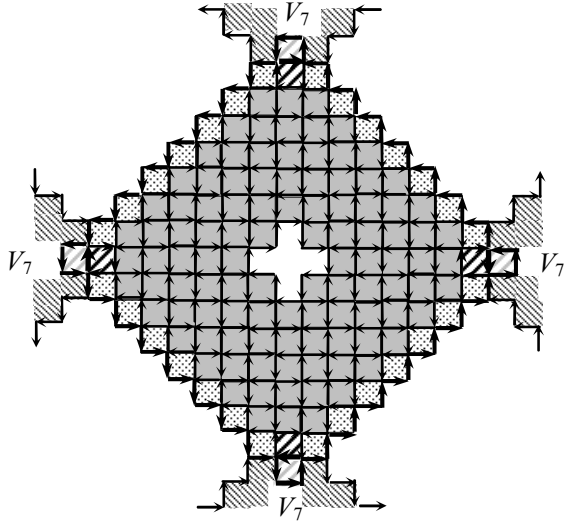

FIG. 6. Sketch of the dynamical $V_8$-$V_7$ cluster percolation. The loop encircled by 68 thick arrows is uncorrected $V_8$-cluster. The corrected number of interfaces of $V_8$-cluster is 60. Each of the 4 small squares in the reference $V_7$-cluster, marked by thick upward diagonal lines, respectively is also the mosaic cell of the corresponding neighboring $V_7$-cluster. The central 5 cavity cells representing the interface tensions of 5 particles will be fully relaxed so as to first jump out of the projection plane after time $t_8$.

$N_8$ (corrected by percolation) = 136                       (5)

Each of the 136 excited particles in eq (5) also connects with a covalent bond in excited state with excited energy $\Delta\varepsilon$. Therefore, the energy of cooperative migration along one direction *in a domain* is $136\Delta\varepsilon$, or $17\varepsilon_0$ (see eq (3)), which comes from the contribution of the specially selected all excited interfaces, as shown in Fig.6, in ideal random distribution. As the migrating direction can be statistically selected as x-, y- z-axial, the *average energy of cooperative migration*, $E_{mig}$, *along one direction in a percolation field* (formed by 2-D clusters in different direction) can be denoted by random thermal energy of $kT_2$, and

$E_{mig} = kT_2 = 17/3\,\varepsilon_0$                            (6)

The random motion energy of $kT_2$ is similar to the energy of Curie temperature in magnetism. It will be proved (in author's next paper: Theoretical proof of the WLF equation) that $kT_2$ here is also the energy of a "critical temperature" existing in the glass transition presumed by Gibbs based on thermodynamics years ago [46]. The denotation of $kT_2$ is the same as that of Gibbs.

The average energy of cooperative migration along one direction, e.g., along the direction of external stress in the glass transition, is an intrinsic attractive potential energy that balances the random motion kinetic energy for flexible polymer system, *independent of temperature and external stress and response time*. That is one of the key concepts to directly prove the WLF equation.

The picture implies that there is a *long-range correlation effect* (with other interfaces in different local fields) of a short-range interaction-interface through its 8 orders of relaxation times and mosaic geometric structure. Especially, the 4 inverted interfaces of the modified 8th loop in Fig.6 denotes that each one of the 4 *inverted slow action torques* still interacts with, e.g., the 1-2 interface in Fig.2 in local 2-D projection plane. Thus, the reference particle $V_0(a_0)$ has not yet *fully* re-coupled with other clusters at local time $t_8$. *I*-th order of reference cluster is always covered with reversed (*i*+1)-th order of reference clusters. The 2-*D* picture is similar to that of the famous critical phase transition based on Kadanoff's [42] study on isotropic ferromagnetism. One important conclusion is that the generating of a $V_8$-cluster always comes from the contributions of its 4 neighboring $V_7$-clusters. That also implies that all neighboring excited local fields in system are simultaneously excited with phase difference of $\pi$ *with* each other, which perhaps is from the period character of ripplons [30].

In the discussion above, the 8 orders of closed loops in Fig.6 are the *track records* of selected excited interfaces, from many random excited interfaces in 3-*D* local space, respectively at the instant of $t_1$, $t_2$…$t_8$ time on a 2-*D* projection plane. In other words, in 3-*D* local space, the cyclic direction, selected from many random oriented interfaces, returns to the direction of ± z-axial at the instant of $t_i$ time, which indicates that the distribution of the effective excited interfaces in 3-*D* local space obeys that of the Graph of Brownian Motion. Therefore, it can be guessed that the fractal dimension of the excited interfaces in the glass transition is 3/2, according to that of Graph of Brownian Motion [47].

### III. 8 ORDERS OF HARD-SPHERES AND 3-D CLUSTERS

#### A. Complex jamming behavior

It is emphasized that the picture of the concomitant 8 orders of transient 2-*D* cluster with 4 interface relaxations of a reference particle only respectively appears at the discrete time $t_i$ in reference local field, the *i*-th order of cluster, being of 5 inner degrees of freedom and one outer degree of freedom, excites cluster possible tiny displacement along ±z-axial during the time of ($t_i$, $t_{i+1}$). For flexible polymer system, it can be noted that if $V_0(a_0)$ with the relaxation time $\tau_0$ takes the z-axial positive direction, the concomitant first order cluster $V_1(a_0)$ with ($\tau \leq \tau_1$) will take the z-axial negative direction, and the concomitant cluster $V_2(a_0)$ with ($\tau \leq \tau_2$) positive, and $V_3(a_0)$ with ($\tau \leq \tau_3$) negative and so on. Therefore, the energy to excite reference particle $V_0(a_0)$ migration along z-axial repeatedly takes $\varepsilon_0(\tau_0)$, -$\varepsilon_0(\tau \leq \tau_1)$, $\varepsilon_0(\tau \leq \tau_2)$, -$\varepsilon_0(\tau \leq \tau_3)$… until $\varepsilon_0(\tau \leq \tau_8)$. During (0, $t_8$), through the excitation of increasing concomitant orders of $V_0(a_0)$ with flipping signs of direction, the reference particle $V_1(a_0)$ migrates laggardly along the z-axial flipping the sign of direction. Since the z-axial excited energy $\varepsilon_0(\tau) = 8 \Delta\varepsilon$ ($\tau \leq \tau_i$) has *i* orders of relaxation times and the interface excited energy quantized, all the migrations of $V_0(a_0)$ along z-axial in fact *cancel out* until the appearance of the largest $V_8(a_0)$, when the 4 interface tensions of $V_0(a_0)$ may be suddenly fully relaxed (see following section) and $V_0(a_0)$ will jump out the bondages (in z-axial or on x-y projection plane) of its 4 neighboring particle fields. This migrating manner of particles and clusters is similar to that of "tube model" in polymer physics. It can be seen that during (0, $t_8$), all particles in local field are in different degrees of relaxation state or jamming state, *none of particles is in fully relaxation state* and solely obtains the evolution energy $\varepsilon_0$ to migrate along z-axial. That further describes the details of jamming behavior in the glass transition.

#### B. Self-similar hard-spheres and 3-D clusters

The conditions of particle $V_0(a_0)$ migrating along the z-axial is that $V_0(a_0)$ should be of fully relaxation interface tensions and 5 degrees of freedom. However, at the instant time $t_8$, the interface tensions of particle $V_0(a_0)$ has not yet been fully relaxed because of the percolation influence: 4 interface tensions with reverse direction, marked by 4 thick arrows in Fig. 6, inlaid on the $V_8$-loop, thus the *reverse torques* with $\tau_7$ react on the interfaces of $V_0(a_0)$ and the particle $V_0(a_0)$ is still to a certain extent astricted from its 4 long-distance *neighboring excited fields* in Fig.6. These anharmonic interface reverse torques are in nature the *long range attractive interactions* between reference $a_0$ local field and other local fields in time and space, which result from the anharmonic vibrations in ripplons.

Nevertheless, at a certain time of $t > t_8$, 4 neighboring first order excited clusters of $V_1(a_0)$, i.e. $V_1(b_0)$, $V_1(c_0)$, $V_1(d_0)$ and $V_1(e_0)$, each containing their 5 particles in respectively 3-D local space, can *dynamically take any directions*, relative to that of $V_1(a_0)$ through fluctuation and one by one *fully eliminate* the anharmonic interface torques between each other, thus the first order of 'hard-sphere' with diameter $\sigma_1$ can be statistically constituted. The first order of hard-sphere $\sigma_1$ contains 5 (first order clusters) × 5 (particles/per first order cluster) = 25 chain-particles that are *compacted* to form a hard-sphere or a 3-D cluster which has an internal density larger than average, because the interface extra volumes (the volume of torque or say, the additional Lindemann distance increment) on $V_0$ inside $V_1$-clusters have been vanished. Note that a concept has been introduced that the interface torque is eliminated accompanied with the extra volume vanished on the interface. This 3-D cluster is a 'dynamical hard-sphere' surrounded by *finite* acting facets that occur *in different local space-time coordinate systems*. The so-called hard-sphere only means that the interaction-interface energy of its two-body is quantized on 2-D local projection plane, independent of the distance of two excited cents in 3-D space. Instant cluster size $\sigma_1 = 25^{1/3} \approx 2.9$ (chain-particle units), is as same as [9], denoting the free motion zone size for a chain-particle. However, specially note that the characteristic size of $\sigma_1$ appears after time $t_8$ is not only correlated with the relaxation time $\tau_1$ and $\sigma_1$ size results from the cooperative contributions of all relaxation times in the local zone $V_1$. That refracts a complex *frequencies implicated effects* of the characteristic size of $\sigma_1$.

It is easy to understand that the length of 25 chain-particles *in free states in z-axial*, instead of in 3-D local space, is the statistical length of first order of segment $l_1$ in flexible macromolecular system.

The remnant interface tensions on first order of hard-sphere can be further relaxed in a larger hard-sphere. The torsional (stress) waves along chain-z-component space direction can be regarded as the parts of ripplons, therefore, the statistic model of 8 orders of self-similar hard-spheres (the self-similar 3-D clusters in small molecular system, or the self-similar segment sizes in macromolecular system), named as $\sigma_i$ (or $l_i$), can be one after another constituted; and each characteristic size of $\sigma_i$ has the frequencies implicated effects that give rise to the complicacy of the glass transition.

The number of particles in the 8th order of hard-sphere, $N(\sigma_8)$, (or the 8th order of segment sizes, $l_8$, in macromolecular system), also is the number of particles $N_c$ in encompassing rearrangements and eliminating interface torques of a reference domain in 3-D space, can be easily found out from Fig.6: the number of cooperative excited particles *along one direction* in corrected $V_8$ is 136 (see II H), the number of interfaces of $V_8$ before percolation is 68, from which subtracts 4 mutual interfaces (that belong to 4 neighboring local fields) of $V_8$ and its 4 neighboring $V_7$-clusters when percolation appears, and each of the 64 (= 68 – 4) interfaces *respectively* relates to an external excited particle in 3-D local space-time coordinate systems, which can eliminate the corresponding interface torque on $V_8(a_0)$ *along z-axial*. Thus, the certain number of particles in the 8th order of hard-sphere $\sigma_8$ is 136 + 64 = 200, a constant for ideal system, i.e.

$$N(\sigma_8) = l_8 = N_c = 200 \tag{7}$$

The numerical value is consistent with the conjectural results of encompassing rearrangements in [40]. Note that $N_8$ in eq (5) represents the number of *cooperative excited particles along z-axial* and $N_c$ in eq (7) is the number of *cooperative rearrangement particles in z-component space*, which lets 136 particles in eq (5) eliminate interface torques *in z-axial* with each other and in free states *in z-axial*.

A foremost result for polymer physics is that the mode of multi-macromolecular motion is that of the slow inverse cascade orientation and the fast cascade re-orientation (relaxation), the degrees of freedom for a macromolecular chain, in free states in 3-D space, is not directly ratio to the chain length $N$. The critical chain length *in free states in z-component space*, instead of 3-D space, for flexible system is 200 (chain-particle units), independent of temperature, which is an inherent character reflecting the motion of macromolecular chain in the range of the temperature from $T_g$ to $T_m$. The numerical value is in accordance with the experimentally determined critical entanglement chain length of ~ 200.

$\sigma_8 = 2\sigma_1 \approx 5.8$ (chain-particle units), which is the size of "cage" in the glass transition. According to the definition of clusters in this paper, an *interaction on* x-y *projection plane* of two *i*-th order of 3-D clusters is always equal to the surface exchanging interaction with $\tau_i$, i.e., the quantized *exchange energy* $\Delta\varepsilon(\tau_i)$, *independent of the distance of two clusters in 3-D space*. That means that the cluster in [17] turns out to be *i*-th order of hard-sphere. During the time of ($t_i$, $t_{i+1}$), *i*-th order of 3-D clusters cannot be welded together, as same as [17], however, at the instant time $t_{i+1}$, all *i*-th order of 3-D clusters in local field will be compacted together to form a compacted (*i*+1)-th order of 3-D cluster with the evolution direction of the first order of 3-D cluster, corresponding to (*i*+1)-th order of local cluster growth phase transition in the glass transition.

### C. The nature of the macromolecular entanglement

The steady excited energy is exactly the critical flow-percolation energy in percolation on a continuum, or say, the mobility edge of classical model in condensed matter physics [45]. The energy of steadily excited state energy flow in the process of $V_8$ vanishing and reoccurring is defined as the localization energy, named as $E_c$ (the same denotation of $E_c$ as Zallen did [45]) in the glass transition. The 8 orders of mosaic geometric structures directly manifest that *the outer degree of freedom of an* (inverse cascade-cascade) *excited state energy flow is 1 and the inner degree of freedom of an i-th order cluster (i < 8) in free state is 5* in the solid-liquid glass transition.

It is a very important theoretic consequence and also foreshows that the outer degrees of freedom of an excited state energy flow is 5 in the melting state phase transition and that for flexible polymer system, the completely renewed energy of a microscopic cluster, i.e. the energy for an *i*-th order cluster (*i* < 8) is in free motion within (*i*+1)-th order cluster zone, is *numerically* equal to the energy of the macroscopic melting state of $kT_m$, namely, $kT_m (\tau \leq \tau_i, i < 8)$.

This also manifest that in tube model, the motion of *i*-th order of segment in tube is not independent. Therefore, the key point in proving the macromolecular entanglement law is to find out the required degrees of freedom, $N^*$, in free states for a chain length $N$ in 3-D space. According to the model of particle-cluster migration along one direction in this paper, an important equation is found out

$$N^* = N_x^* \cdot N_y^* \cdot N_z^* = (N_z^*)^3 \tag{8}$$

in which $N_x^* = N_y^* = N_z^*$, and $N_z^*$ is the required degrees of freedom in z-axial thawing for a chain length $N$. Namely, de Gennes has only theoretically deduced the relationship of $\eta_z \sim (N_z^*)^3$ in [44] for a *free chain* $N_z^*$ *in z-space* $\in d^3$, instead of chain $N$ itself in 3-D space. Therefore, formula (9) is deduced

$$\eta \sim (N_z^*)^9 \tag{9}$$

Since the molecular weight $M \sim N_z$, $N_z$ is the z-component of chain length $N$, the key point in proving the macromolecular entanglement law thus turns to finding out the scaling relationship between $N_z$ and $N_z^*$. When $N_z > 200$ (that provide all 320 z-component excited interface states), it is certain that the z-component interfaces of some chain-particles in a macromolecular chain are in unexcited states and the motions of these chain-particles are frozen in z-component space, even though the macromolecular chain has been in melt state. However, the mobility edge energy $E_c$ of the directional thaw of macromolecular chain and the random motion energy $kT_m$ are independent of chain length (for $N > 200$). These chain-particles in freezing motions are exactly the so-called entanglement structure. This theoretic consequence foreshows that the ratio of $E_c/kT_m$, reflecting the balance between directional and random motions in system, is sure to appear in the exponential expression of $M$. (According to the result, it will be directly proved $\eta \sim M\exp\{9(1-E_c/kT_m)\}$ for really macromolecular system and $\eta \sim M^{27/8} = M^{3.375} \approx M^{3.4}$ for flexible polymer system, as shown in author's next paper.)

### D. Localization energy $E_c$ in the glass transition

Disorder-induced localization is the essential concept of Anderson's disorder theorem in condensed matter physics [45]. It can be seen from Fig.6 that the localization object induced by thermo-disorder in the glass transition is just the 8th order symmetric loop in interface tension wave or the *localization energy* to excite the 8th order symmetric loop. The localization here also can be looked upon the 'cage effect' in the glass transition. Now the geometric method is used to derive the localized energy, $E_c$, induced by thermo-disorder in the glass transition. $E_c$, can be also called the critical flow-percolation energy in percolation theory.

There are two approaches to derive $E_c$ according to the mean field theory of mosaic structure. One takes the dimension of cage as the largest closed cycle, and derives $E_c$ through the structure of the self-similar 2-D closed cycles from large to small. The other takes the dimension of cage as the smallest 5- particle (i.e., $a_0$, $b_0$, $c_0$, $d_0$, $e_0$ in Fig.3-6) excited field and derives $E_c$ through the *random motion structure* of self-similar 5-particle excited fields from small to large.

The step (interface) number of the 8th order 2-D loop is $L_8 = 60$ (see II H). The energy of all steps is $60\Delta\varepsilon(\tau_8)$ which is derived from the geometric structure of a lone local field in Fig.6. The occurrence of 60 steps in a $V_8$-loops is necessary in forming percolation, however, the critical energy forming flow-percolation, $E_c$, is less than $60\Delta\varepsilon(\tau_8)$ because of the quantized energy effect of excited interface and the dynamical mosaic structure of excited interface energy flow. The dynamical continuum energy of an excited state energy flow, in Fig.6, can be used to deduce the flow-percolation energy $E_c$. The interface excited energies on $V_8$-loop in Fig.6 are *shared* by $V_0(\tau_0)$-$V_8(\tau_8)$ interfaces and $V_7(\tau_7)$-$V_8(\tau_8)$ interfaces. In inverse cascade, an excited interface with energy of $\Delta\varepsilon(\tau_8)$ has "connected" with many excited interfaces with relaxation time of $\tau_i < \tau_8$, in large area. Contrarily, the energies of $V_i$-loops with relaxation time of $\tau_i < \tau_8$ also can form a few of the interfaces of a $V_8$-loop. The key concepts are that inverse cascade does not dissipate excited energy and all the excited energies *forming* a $V_8$-loop in Fig.6 can not be *simultaneity* obtained at a certain observation time in the reference $a_0$ local field, a part of which comes from the inverse cascade energies that is of all *i*-th order of loops in the neighboring local fields. This is the dynamical mosaic structure of excited interface energy flow. The flow-percolation in the glass transition is just the continuum of excited state energy flow formed from $V_0(\tau_0)$ loops → $V_8(\tau_8)$ loops in system. In other words, a few of interfaces (named as $L_{\text{inverse}}$) on the $V_8$-loop in a reference $a_0$ local field will be, in manner of slow inverse cascade, excited by the interfaces with relaxation time of $\tau_i < \tau_8$ in the new local fields after $a_0$ (mosaic structure!), and the others (named as $L_{\text{cascade}}$) on the $V_8$-loop will one by one vanish and their excited energy with relaxation time of $\tau_8$ will rebuild many new $\tau_0$-interfaces, in a manner of *fast cascade vibration* (seeing the next paper of 'Fixed point for self-similar Lennard-Jones potentials in the glass transition'), in the new local fields. Thus, in the flow-percolation on a continuum, the 60 interfaces of a reference $V_8$-loop in Fig.6 are *dynamically* divided into two parts: the $L_{\text{cascade}}$ interfaces that occur at the local time of $t_8$ in $a_0$ field and the $L_{\text{inverse}}$ interfaces that are the mosaic structure of energy flow and occur at a time after $t_8$. Formula (10) is obtained

$$L_{\text{cascade}} = L_8 - L_{\text{inverse}} \tag{10}$$

The energy of $L_{\text{cascade}}$ is also the slow-process inverse cascade energy that changes as fast-process cascade vibrant energy of rebuilding new $V_0$-loops when the $L_{cascade}$ interfaces vanish. The energy of $L_{inverse}$ is the slow-process inverse cascade energy of all $V_i$ loops from $V_0$ to $V_7$ in new local fields. The dynamical balance in flow-percolation is that the energy of $L_{cascade}$ interfaces equals to that of $L_{inverse}$ interfaces. The balance excited energy of inverse cascade-cascade in flow-percolation is exactly the localize energy $E_c$ in the glass transition.

Therefore, $E_c/\Delta\varepsilon(\tau_8) = L_{cascade}$ in which $\tau_8$ is used as timescale of $V_8$ vanishing. Since the excited energy in inverse cascade is not dissipated, the evolution energy of each order closed cycle is $8\Delta\varepsilon(\tau_i)$ $=\varepsilon_0(\tau_i)$, the singular point energy of $i$-th order closed cycle. The singular point energy of the closed cycle of forming 60 steps is $\varepsilon_0(\tau_8)$. Therefore, the average number of $L_{inverse}$ in eq (10) is $E_c/8\Delta\varepsilon(\tau_8)$. The energy of each step on a $V_8$-loop is still $\Delta\varepsilon(\tau_8)$ and the step number of the interfaces of a $V_8$-loop is still 60. From eq (10), formula (11) is obtained

$$E_c = 60\Delta\varepsilon(\tau_8) - E_c/8 \tag{11}$$

From (11), formula (12) is obtained

$$E_c = 320\Delta\varepsilon/6 = 20/3\,\varepsilon_0 \tag{12}$$

By recursively using eq (11), the following is obtained

$$E_c = 60\,\Delta\varepsilon - 1/8\,(60\,\Delta\varepsilon - 1/8\,(60\,\Delta\varepsilon - 1/8\,(60\,\Delta\varepsilon - \ldots))) \tag{13}$$

$$= 60\,\Delta\varepsilon\,(1 - 1/8 + (1/8)^2 - (1/8)^3 + \ldots) = 60\,\Delta\varepsilon\,(8/9) = 20/3\,\varepsilon_0$$

Eq (13) also gives an expression to the localizability of $i$-th order $2\pi$ (closed cycle) energy flow with relaxation time $\tau_i$ in inverse cascade. Thus, eq (12) can be rewritten as

$$E_c(\tau_i) = 320\Delta\varepsilon(\tau_i)/6 = 20/3\,\varepsilon_0(\tau_i) \tag{14}$$

Eq (14) denotes that the localized energy $E_c$ (the mobility edge energy, the critical flow-percolation energy) in the glass transition has 8 components, $E_c(\tau_i)$, $i = 1, 2\ldots 8$ and the numerical value of each component energy is the same, namely, $20/3\,\varepsilon_0$. This is one of the singularities in the glass transition.

On the other hand, the topological geometric structure in percolation field can be simplified as self-similar 5-particle excited field in which cage is the smallest unit. Each cage is a 1-st order 5-particle excited field with 5 degrees of freedom. Every 5 cages can form a 2-nd order 5-particle excited field and every 5 2-nd order 5-particle excited fields form a 3-rd order 5-particle excited field... The energy to form a reference cage is the inside field energy of the cage, $5\varepsilon_0$, plus the contribution from 4 one by one outside fields, thus

$$E_c = 5\varepsilon_0 + E_c/4 = 5\varepsilon_0 + 1/4\,(5\varepsilon_0 + 1/4\,(5\varepsilon_0 + 1/4\,(5\varepsilon_0 + \ldots)))$$

$$= 5\varepsilon_0\,(1 + 1/4 + (1/4)^2 + (1/4)^3 + \ldots) = 5\varepsilon_0\,(4/3) = 20/3\,\varepsilon_0 \tag{15}$$

The $E_c$ on both sides of eq (15) indicate that the random energy is the same both inside and outside field (mean field theory). The 4 neighboring cages outside the reference cage are interlaced in time and take turn to interact with the reference cage. Therefore, eq (15) reflects a random interface excited structure of glass transition in which *the outside field always contributes to the inside field*.

The results of eqs (12)- (15) indicate that the localized energy induced by thermo-disorder, $E_c(\tau_i)$, is a characteristic invariable with 8 order of relaxation time spectrum in the glass transition, independent of glass transition temperature $T_g$, which, comes from the mean filed of eq (13) or eq (15) being of mosaic structures, may be of minimum energy to excite glass transition. The localized energy $E_c(\tau_i)$ is very important to interpret the macroscopic glass transition. If an *thermo random motion energy*, $kT_g°(\tau_i)$, of *i*-th order clusters with relaxation time $\tau_i$ is used to denote the energy of $E_c(\tau_i)$, that is,

$$kT_g°(\tau_i) = E_c(\tau_i) = 20/3\,\varepsilon_0(\tau_i) \tag{16}$$

The numerical value of $T_g°(\tau_i) = T_g°$ that can be called the *fixed point* in renormalization of clusters from small to large, in the critical local cluster grow phase transition in glass transition, it is independent of glass transition temperature $T_g$. Furthermore, in the author's next paper 'Fixed point for the second Virial coefficients in the glass Transition', the statistical physics will be used to prove the validity of formula (16) independently and the formula $T_g°(\tau_8) = T_g$ will also be introduced. (Here, $T_g$ is traditionally accepted as glass transition temperature, which is obtained by slow heating rate) Therefore, $E_c$ is a measurable magnitude by experiments. From eq (6), numerical relationship is:

$$kT_g° = kT_2 + \varepsilon_0 \tag{17}$$

It can be seen that the eq (12) is based on the result of corrected $L_8$ value, and this correction from $L_7$ to $L_8$ in Fig.6 also can be regard as the geometric frustration effect, which appearance corresponds to the glass transition [14].

### E. The number of excited states and activation energy to break solid lattice

Fig.6 has given out the number of all excited states in the glass transition: the 320 different interface tension relaxation states. The numerical value is from

$$\sum_{i=1}^{8} L_i + 8 = 320 \tag{18}$$

The $L_i$ is the number of the interfaces of *i*-th order 2-*D* loop. The numerical value of 8 in eq (18) is the 8 evolution states, i.e., the difference between the uncorrected and the corrected numbers of interfaces of 8th loop. The 8 evolution states together with the 4 fully relaxed interface states on a reference particle $a_0$ will evolve a new first order asymmetry loop being of 12 interfaces when 8th cluster appears. It can be seen that a mode of the glass transition has been proposed, which is also the mode of breaking solid lattice. For flexible polymer system, in brief, the mode to break a 3-*D* solid domain is the interface-torque progressive relaxation of the 8 orders of 2-*D* clusters in local z-space, which relates to the 320 different interface tension relaxation states. The energy summation of the 320 different interface tension relaxation states is named the *activation energy* to break solid lattice, denoted as $\Delta E_{co}$. (The reason to call it activation energy is that although in macroscopically, the energy to break solid lattice is only $kT_g$, in microcosmic, 320 interface energy of every solid domain is needed. From eqs (3), (12), (17), (18), equation (19) is obtained

$$\Delta E_{co} = 320\Delta\varepsilon = 40\varepsilon_0 \tag{19}$$

And 
$$E_c = kT_g° = \frac{1}{6}\Delta E_{co} \tag{20}$$

Using $T_g°(\tau_8) = T_g$, thus,

$$kT_g = \frac{1}{6}\Delta E_{co} \tag{21}$$

The ratio of $E_c/\Delta E_{co} \equiv \phi_c(E_c) \equiv 1/6$, its physical meaning is that $\phi_c(E_c)$ specifies the occupied fraction of 320 interface-excited states that allow flow of energy $E_c$ to occupy. The ratio is consistent with the result of Zallen [45], who suggests $\phi_c(E_c) \approx 0.16$ in 3-*D* space. Here, the occupied fraction of all interface-excited states allowed to flow of energy $E_c$, takes the place of the occupied fraction of space allowed to particles of energy $E_c$ proposed by Zallen.

An important conjecture comes from eq (20), if the average thermo random energy $kT = kT_g^\circ(\tau_8) = kT_g$, the glass transition would occur after time $t > \tau_8$, while if $kT < kT_g^\circ(\tau_8)$, the lower temperature glass transition also can occur, depends on the thermo-fluctuation and takes long time to obtain the occupied fraction $\phi_c(E_c)$ in a domain. The occupied fraction is also an invariant, independent of the glass transition temperature, which is also one of the singularities in the glass transition. The increase of transition temperature only increases the number of excited interface energy flow, or the number of thawing domains to expedite glass transition.

As the mode of breaking solid lattice is also that of the cooperative migration to excite particle-clusters along one direction, it is presupposed that $\Delta E_{co}$ the activation energy to break solid lattice is also the activation energy of *cooperative orientation* of macromolecules in elongation flow. It is interesting that the concept of activation energy is widely applied in polymer processing. Here, the orientation activation energy in polymer physics is assured to come of all the interface excited states in system. As the method of selected 320 interface states along one direction in ideal random system obeys that of the Graph of Brownian Motion, the relationship between stretch-orientation viscosity and $\Delta E_{co}$ obeys $\eta_{co} \sim \exp(\Delta E_{co}/T)$. Therefore, $\Delta E_{co}$, $\Delta \varepsilon$ and $\varepsilon_0$ (eq (19)) can also be measured by the stretch-orientation experiment with melt high-speed spinning along one direction. In fact, the author [48] has obtained the experimental value of orientation activation energy $\Delta E_{co}$ for polyethylene terephthalate (PET), by using the on-line measuring on the stretch-orientation zone during melt high-speed spinning at 2200–4200M/min. The experimental result shows $kT_g$ (for PET) ≈ 1/6 $\Delta E_{co}$ (for PET), satisfying eq (21), and $\Delta E_{co}$ (for PET) = 2035$k$. Thus, $\Delta \varepsilon$ (for PET) = 6.4 $k$ = 0.55meV, and $\varepsilon_0$ (for PET) = 51$k$. (Furthermore, it will be proved that $\varepsilon_0/k$ is exactly the constant $C_2$ = 51.6 in WLF equation)

### F. Free volume

The classical free volume ideas [41] have been questioned because of the misfits with pressure effects, but here the pressure effects should be indeed minor [17]. It is interesting in our model that there is no so-called classical free volume with an atom migrating in system unless percolation appears. Except the 8th order of 2-*D* clusters (hard-spheres), in the *i*-th order of local phase transition, all (*i*-1)-th order of 2-*D* clusters are compacted to form *i*-th order of 2-*D* clusters, and the extra volumes of the surfaces of all (*i*-1)-th order of 2-*D* clusters vanish and reappear on the surfaces of *i*-th order of clusters. So it is proposed that when the 8th order of local cluster growth phase transition (i.e., percolation) appears, the extra volumes in the surfaces of the 5 7-th order of 2-*D* clusters suddenly form vacancy volumes of 5 cavity sites in Fig.6. Here the definition of free volume is modified by the cooperatively appearing of 5 cavity sites, i.e., using clusters rather than atoms, as same as that proposed by de Gennes [17]. The modified free volume ideas is still true, because the action of pressure should be only through the same percolation field generating cavity volume and the cavity volume composed by extra volumes in ripplons is of interface tension energy, which should be balanced with the external pressure or tension. As the average occupying cavity volume per particle (involved its all states in cooperative rearrangement) is as the free volume fraction, and the 5 cavity volumes only occur during a cycle period in structure rearrangements, which need 200 cooperative rearrangement particles in local z-space, thus the free volume fraction: 5/200 = 0.025 can be directly and explicitly obtained, which is in accordance with the experimental results for flexible polymer (so-called the free volume theory in the glass transition).

### IV. CONCLUSION

This paper is also a theoretical preparative to directly prove the two well-known empirical laws, WLF equation and the macromolecular entanglement law. In order to research into the universal picture of particle-clusters cooperative migration in one direction and the minimum energy to excite glass transition, the intrinsic 8 orders of 2-*D* mosaic geometric structures has been proposed by geometric method. The so-called dynamical behavior may actually means that only when backmost in all in-order arisen interface states on 8 orders of 2-*D* structures is excited, can the glass transition appear. The central assumption is that the additional restoring force moments on an excited interface has and only has 8 orders of relaxation times. Although the assumption in the paper is based on Wolynes' the limited domain wall vibration frequencies, it will be independently proved in the author's other paper 'Fixed point for self-similar Lennard-Jones potentials in the glass transition'. It will be further proved that there are only 8 orders of self-similar hard spheres in the glass transition. One of the most important results is that this model may well provide a unified mechanism to interpret hard-sphere, compacting cluster and cluster growth, density fluctuation, free volume, cage, geometric frustration, jamming behaviors, Ising model, reptation, percolation, potential energy landscape, breaking solid lattice, cooperative migration, cooperative orientation, critical entanglement chain length and structure rearrangements. More important, for flexible polymer system, it is the intrinsic 8 orders of mosaic structures that directly deduce a series of quantitative values: the average energy of cooperative migration along one direction $E_{mig} = kT_2 = 17/3\varepsilon_0$, the number of cooperative rearrangement particles in z-component space, also is the critical entanglement chain length, $N_c = 200$, the flow-percolation energy or the localized energy, $E_c(\tau_i) = kT_g^\circ(\tau_i) = 20/3\varepsilon_0(\tau_i)$, the activation energy to break solid lattice, $\Delta E_{co} = 40\varepsilon_0$.

In flexible polymer system, all 320 different interface excited states in turn occur on 8 orders of 2-*D* mosaic lattices and all the excited states possess the same numerical value of quantized interface excited energy $\Delta\varepsilon \approx 6.4k \approx 0.55$ meV, but have different interacting times, relaxation times and different time-space phases. Furthermore, the dynamical rule of these excited states should, from fast to slow and small to large, form instant 8 orders of $2\pi$ interface excitation loops in inverse energy cascade, with the even number cycles along one direction and the odd number cycles, the opposite direction.

The picture of particle-clusters cooperative migration in the ideal glass transition can be simply described as that in a reference particle $a_0$ z-component space, the 320 interface excitations, appearing one by one, on the 8 orders of 2-*D* mosaic lattices formed by 200 particles in z-component, step by step and from fast to slow induce 141 particles to move back and forth along local ±z-axial. Once 320-th interface excitation appears, the +z-axial $a_0$ particle immediately migrates just one particle-distance away from its 4 neighboring -z-axial particles and gives rise to 5 z-axial particle-cavities. For the subsequent $a_0$ particle in the other arbitrary z'-component space, the same process will occur. Only after a long time and repetitious rearrangements has the reference $a_0$ particle full re-coupled with its primal 4 neighboring particles.

In multiparticle system, it is very difficult to obtain such discussion result as the 8 orders of 2-*D* mosaic lattices independently from any single coordinate system either starting from a random Hamiltonian or the fully microscopic mode-coupling theory. Therefore, the geometric method, (the standout mode of an arrow on an excited interface, also connects 8 orders of coupling clusters, similar to that of mode-coupling theory, contains much of topological characters in Fig.6, especially, Fig.6 is similar to that of classical hydrodynamics, which refracts the crossover from glass transition to liquid state and will be discussed in next papers) is used to find out the minimum energy to excite the glass transition. In the geometric method, only a set of discrete time $t_i$ in a reference local field is modally used to represent the occurrence of transient *i*-th order 2-*D* cluster. It is stressed that when the reference local coordinate system occurs at time $t_i$ its local coordinate system at time $t_{i-1}$ has vanished in dynamic evolution of local field.

The author's next paper: 'Fixed point for self-similar Lennard-Jones potentials in the glass transition' will prove that the origin and transfer of interface excitation come of the balance effect between self-similar Lennard-Jones potential fluctuation and geometric phase potential fluctuation.

The theoretical framework of ideal glass transition in the paper has provided a foundation for the further research on the glass transition and also offered a new way to study glass transition of complex system.

This paper greatly benefited from comments and criticisms by Yasmin McGlashan and Beatrice Greaves. The author is grateful to Professor Yun Huang of Beijing University, Professor Da-Cheng Wu of Sichuan University, Professor Bo-Ren Liang and Professor Cheng-Xun Wu of Donghua University for valuable and enlightening discussions.